\documentclass[preprint2]{aastex}

\usepackage[usenames]{color}    

\slugcomment{}

\shorttitle{Origin of plasma in the 2011 June 7 filament}
\shortauthors{Yardley et al.}

\begin{document}

\title{Flux cancellation and the evolution of the eruptive filament of 2011 June 7}

\author{S. L. Yardley\altaffilmark{1}, L. M. Green\altaffilmark{1}, D. R. Williams\altaffilmark{1}, L. van Driel-Gesztelyi\altaffilmark{1,2,3}, G. Valori\altaffilmark{1} \& S. Dacie\altaffilmark{1} }
\affil{\altaffilmark{1} Mullard Space Science Laboratory, University College London, Holmbury St. Mary, Dorking, Surrey, RH5 6NT, UK}
\affil{\altaffilmark{2} Observatoire de Paris, LESIA, UMR 8109 (CNRS), F-92195 Meudon-Principal Cedex, France}
\affil{\altaffilmark{3} Konkoly Observatory of the Hungarian Academy of Sciences, Budapest, Hungary}
\email{stephanie.yardley.13@ucl.ac.uk}

\begin{abstract}

We investigate whether flux cancellation is responsible for the formation of a very massive filament resulting in the spectacular 2011 June 7 eruption. We analyse and quantify the amount of flux cancellation that occurs in NOAA AR 11226 and its two neighbouring ARs (11227 \& 11233) using line-of-sight magnetograms from the Heliospheric Magnetic Imager. During a 3.6-day period building up to the filament eruption, $1.7 \times 10^{21}$~Mx, 21\% of AR 11226's maximum magnetic flux, was cancelled along the polarity inversion line (PIL) where the filament formed. If the flux cancellation continued at the same rate up until the eruption then up to $2.8 \times 10^{21}$~Mx (34~\% of the AR flux) may have been built into the magnetic configuration that contains the filament plasma. The large flux cancellation rate is due to an unusual motion of the positive polarity sunspot, which splits, with the largest section moving rapidly towards the PIL. This motion compresses the negative polarity and leads to the formation of an orphan penumbra where one end of the filament is rooted. Dense plasma threads above the orphan penumbra build into the filament, extending its length, and presumably injecting material into it.  We conclude that the exceptionally strong flux cancellation in AR 11226 played a significant role in the formation of its unusually massive filament. In addition, the presence and coherent evolution of bald patches in the vector magnetic field along the PIL suggests that the magnetic field configuration supporting the filament material is that of a flux rope.

\end{abstract}

\keywords{Sun: filaments, prominences --- Sun: coronal mass ejections (CMEs) --- Sun: photosphere ---  Sun: activity --- Sun: magnetic topology --- Sun: evolution }

\section{Introduction}
 
Filaments, composed of relatively cool and dense plasma, are a common feature in the solar atmosphere and understanding their formation is an open question in solar physics. Filaments typically take up to several days to form and lie above photospheric polarity inversion lines (PIL) \citep{Babcock-1955}. Despite decades of observations, the conditions for filament formation remain unclear, with a variety of mechanisms proposed to explain the formation process. 
 
To resolve the unanswered question of filament formation, we must consider two aspects. First, the magnetic configuration that can support the plasma in the solar atmosphere against gravity. Regarding the magnetic models the magnetic field configurations do not necessarily invoke dips in the magnetic field e.g. \citet{Martin-1994}. Although, one common feature of filament models is the presence of sheared magnetic field along the PIL. This can either be in the form of a weakly twisted magnetic flux rope \citep{Kuperus-1974, Pneuman-1983, Ballegooijen-1989} or a sheared arcade \citep{Antiochos-1994, Devore-2000, Aulanier-2002}. The second aspect to be considered is the origin of the filament plasma itself. This must be supplied either from the chromosphere, through the emergence of new magnetic flux lifting plasma into the atmosphere \citep{Rust-1994, Deng-2000} or through direct injection of chromospheric plasma \citep{Poland-1986, Wang-1999}, or via condensation from the corona \citep{Engvold-1977, An-1985}.

In this study we investigate the role that flux cancellation may play in the origin of the plasma in an eruptive filament. We also investigate how the magnetic field configuration evolves as flux cancellation is proceeding to find indications of the onset of the filament's eruption as a coronal mass ejection.

Photospheric flux cancellation is observed as the convergence, collision and subsequent disappearance of small-scale opposite-polarity magnetic features in line-of-sight (LoS) magnetograms \citep{Martin-1985}. These cancelling features can be as small as a few hundred kilometres across with magnetic fluxes as low as $\sim10^{17}$~Mx \citep{Litvinenko-1999}. The cancellation process is observed throughout the quiet Sun and in active regions (ARs) at all stages of their evolution. In ARs, the cancellation is observed both between the main magnetic polarities (along the internal PIL) and between magnetic flux at the AR periphery and surrounding magnetic fields.

The opposite polarity fragments that collide and subsequently disappear are interpreted as representing the footpoints of two magnetic flux systems that are sheared across the PIL. These features, during collision, undergo magnetic reconnection, which takes place low in the solar atmosphere; either in the photosphere \citep{Yur-2001, Bellot-2005} or the chromosphere \citep{Litvinenko-1999, Chae-2001, Kim-2001, Litvinenko-2015}.

The magnetic reconnection leads to the formation of two magnetic flux systems that differ in connectivity to the pre-reconnection pair: there is now a small loop with a radius of curvature that provides a strong downward tension force and, above this, the formation of highly sheared field that has a concave-up section above the reconnection region. The disappearance of the opposite polarity fragments during flux cancellation is then the observational manifestation of the submergence of the small loop, whilst the longer loop remains in the solar atmosphere. 

This physical interpretation of flux cancellation is of direct relevance to the study of filaments. During reconnection, the magnetic field diffuses through the dense plasma of the lower atmosphere and theoretical studies have shown that reconnection occurring low in the solar atmosphere can effectively drive the mass required for filament formation upwards \citep{Priest-1996, Litvinenko-1999, Litvinenko-2007}. Therefore, flux cancellation may play a key role in the origin of filament plasma. If this is indeed the case, the quantity of plasma injected into a forming filament could be expected to scale with the amount of magnetic flux cancellation observed.

The physical processes that lead to the observation of flux cancellation can also be used to investigate the magnetic field environment of the filament. At the internal PIL of ARs, the flux cancellation process initially builds highly sheared field along the PIL and dips in this magnetic configuration that could support filament material. However, ongoing flux cancellation by sustained magnetic reconnection will eventually form helical magnetic fields around the sheared arcade. This can therefore reconfigure the magnetic field from a sheared coronal arcade into a magnetic flux rope configuration (see Figure~1 of \citealt{Ballegooijen-1989}). Concave-up sections in a magnetic flux rope configuration also provide locations where filament plasma can be supported. Such a scenario for flux rope formation is well supported by simulations \citep{Amari-2003, Aulanier-2010}. Whether the magnetic field environment of a filament is that of a sheared arcade or weakly twisted flux rope is a challenging question to answer without direct measurements of the magnetic field above the photosphere. However, several observational proxies have been developed. The main proxy relevant to active region filaments are so-called ``bald-patches'' (BPs), where the photospheric vector magnetic field at the underside of the filament is tangent to the photosphere and crosses the PIL in the inverse direction. This is indicative of the presence of concave-up sections of magnetic field, formed, e.g. at the bottom of a low-lying flux rope \citep{Athay-1983, Lites-2005, Lopez-2006, Canou-2009}. Understanding the magnetic field configuration of a filament at its point of eruption is vital for understanding the physical mechanisms the trigger and drive the event. If filaments form in a magnetic flux rope configuration, their eruption as a coronal mass ejection (CME) can be understood as a loss of equilibrium or an ideal instability of the rope \citep{Forbes-1991, Fan-2003, Torok-2003, Demoulin-2010}. If filament material is supported in a sheared arcade the expansion and reconnection within the structure can produce the eruption \citep{Antiochos-1999, Moore-2001}. Flux cancellation observations provide a way to investigate how much magnetic flux has been built into the magnetic configuration before eruption, such as the magnetic flux of the structure containing the filament material in relation to the overlying, restraining field of the AR. This is key to understanding the onset of CMEs.

Previous flux cancellation studies in ARs which form filaments and produce CMEs have exhibited cancellation rates of 10\% of the AR flux per day with a total of $1 \times 10^{21}$~Mx cancelled during the time period studied \citep{Sterling-2010, Baker-2012}. \citet{Green-2011} studied flux cancellation along the PIL of AR 10977 through the entire period commencing at the start of the flux emergence phase, through the decay phase to the occurrence of the first CME produced by the region. They found that $0.71 \times 10^{21}$~Mx ($\sim$34\% of the AR flux) cancelled in the 2.5 days leading up to the eruption.

The study presented here focuses on the flux cancellation that occurs along the internal PIL of AR 11226, during the time that a filament along the PIL is growing in size. The filament erupted on 2011~June~7 at $\sim$06:15~UT. Prior to eruption the filament, located in the southern hemisphere, appeared ordinary in both appearance and size. However, the filament exploded and unleashed a vast amount of material that experienced a huge lateral expansion as it was launched into the solar system. Much of the material fell back towards the photosphere as discrete, dense blobs impacting almost a quarter of the solar surface.

Recent work by \citet{Carlyle-2014} indicates that the filament contained a huge amount of mass, given that, even after expansion, the column densities of the individual blobs are comparable to that of a typical pre-eruption filament \citep{Gilbert-2005}. \citet{Lidia-2014} noticed remarkably strong flux cancellation along the PIL in the four days leading up to the eruption, but the cancellation was never quantified.

The combination of these observations, measurements and the remarkable nature of the erupted material suggests the amount of magnetic flux built into the magnetic configuration of the filament and the free energy stored in the field must have been colossal. We investigate the possibility that the filament that erupted on 2011~June~7 was exceptionally massive because of an especially high flux cancellation rate.

This paper is organised as follows. In Section~\ref{sec:n2} we describe the instrumentation used and the application of the algorithm. In Section~\ref{sec:n3} we present the filament, LoS and vector field evolution, penumbra formation and flux cancellation rate. In Section~\ref{sec:n4} we discuss the unusual photospheric motions and high flux cancellation rate, which is followed by our conclusions in Section~\ref{sec:n5}.

\section{Instrumentation \& Algorithm Application} \label{sec:n2}

\subsection{Instrumentation} 

In this study we compared and analysed data from a wide range of instruments that collectively observe from the photosphere to the corona. The Atmospheric Imaging Assembly (AIA; \citealt{Lemen-2012}) instrument onboard the \textit{Solar Dynamics Observatory} (SDO) provides full-disk observations in three UV continuum wavelengths as well as seven EUV bandpasses with a high temporal and spatial resolution of 12~s and 1.5'' respectively. We focus on the wavebands 304 and 193~{\AA}, which are dominated by plasma emission at temperatures of approximately 0.05 (304~{\AA}), 1.2 and 20~MK (193~{\AA}), to study the location and evolution of the filament material. H$\alpha$ images from the Kanzelh{\"o}he Observatory were also analysed to determine the location of filament material.

The evolution of the photospheric magnetic field is studied using data from the Helioseismic and Magnetic Imager (HMI; \citealt{Schou-2012, Scherrer-2012}) onboard SDO. This includes the calculation of the magnetic flux cancellation rate by using full-disk line-of-sight (LoS) magnetograms computed from filtergrams sampled at six points across the Fe I 6173~{\AA} absorption spectral line. These filtergrams are recorded by the vector field camera with a pixel size of 0.5''~pixel$^{-1}$ and a noise level of 10~G. Multiple measurements are combined to give a 720~s cadence. The CEA HMI SHARP 720s data series \citep{Hoeksema-2014, Bobra-2014} are analysed to investigate the orientation of the transverse component of the vector magnetic field along the PIL.

HMI continuum data is used to study the sunspot evolution and penumbral dynamics during the filament formation. To quantify sunspot proper motions and velocities the Debrecen Photoheliographic Data sunspot catalogue (DPD; \citealt{Gyori-2011}) was used. This catalogue is mainly composed of full-disk white light observations taken at the Debrecen Observatory, although several ground- and space-based observations are now included. It contains accounts of position and area for all sunspots, irrespective of size, with a mean precision of $\sim$0.1$^{\circ}$ and $\sim$10~\% respectively. We identified the main spots and determined the velocity of their proper motion from the published daily positions.

Additionally, 195~{\AA} EUV images produced by the Sun Earth Connection Coronal and Heliospheric Investigation (SECCHI) onboard the \textit{Solar Terrestrial Relations Observatory} (STEREO; \citealt{Howard-2008}) spacecraft are used to observe the emergence of AR 11226, which took place on the far-side of the Sun. The observations are provided by the STEREO-B spacecraft, which is positioned at 93.6$^{\circ}$ away from the Sun-Earth line at the time of the observations.

\subsection{Application of the STEF Algorithm}

The photospheric field evolution of AR 11226, and its two neighbouring regions as a comparison, were studied using the 720~s data series available from the Helioseismic and Magnetic Imager (HMI). The magnetic complexity is monitored, whilst the total flux content of the ARs is quantified by using the Solar Tracking of the Evolution of photospheric Flux (STEF) algorithm. STEF automatically detects and tracks both small- and large-scale magnetic features in LoS magnetograms and is used to study the magnetic field evolution of ARs throughout their lifetimes from flux emergence to dispersal.

AR areas are identified by eye and the field of view is assigned as a rectangular area, the size of which is specified by the user. The radial component of the magnetic field is then estimated by applying a cosine correction using the Heliocentric Earth Equatorial Coordinate system (HEEQ) \citep{Thompson-2006}. The magnetogram containing the radialised field values is then differentially rotated to the central meridian passage time of the region of interest to correct for projection effects using a routine that has been developed in Sunpy.

The flux-weighted central coordinates of this area are calculated at each time step to track the feature such that it remains in the centre of the field of view. The magnetic features are then selected as follows. First a Gaussian filter is used to smooth the data with a standard deviation (width) of the applied Gaussian of 7 pixel units. The weighted average of the magnetic flux density of the neighbouring pixels must exceed a cut-off of 40 G. The largest regions identified are kept and make up at least 60\% of the selected area, whereas the smaller features at large distances are disregarded. This is to remove quiet Sun features which are not part of the AR, although small-scale features will still enter or exit the boundary which introduces a contribution or reduction to the magnetic flux. These fluctuations are at least three orders of magnitude less than the total AR flux.

When used on data from the quiet Sun only, the algorithm applies no smoothing and selects pixels above a threshold of 3~$\sigma$, where $\sigma$ is the accuracy of the LoS magnetic field (e.g. 30~G for HMI and 60~G for MDI). As false positives are more likely to occur in the detection of a small collection of pixels a second criterion is applied, namely that features must also be equal to or larger than 4 pixels in size (0.54 Mm$^{2}$ for HMI, 8 Mm$^{2}$ for MDI).

Once the pixels have been selected, the corresponding magnetic flux density values are extracted, summed and multiplied by the area to obtain the total magnetic flux. The outputs of this algorithm include: total positive and negative pixel area; total positive, negative and unsigned magnetic flux; the distance and tilt angle between the flux-weighted central coordinates. Kernel density estimation plots are created, using either a Gaussian or a box kernel to show the frequency of pixels with respect to flux density.

The amount of flux cancelled is calculated from the reduction in total magnetic flux. This approach can be used in this AR since flux cancellation at the AR periphery and flux fragments leaving the boundary of the region are are at least three orders of magnitude less than the total AR flux. STEF calculates both the positive and negative flux of the AR but we focus on the total unsigned magnetic flux (half the total positive and negative flux) for calculating flux cancellation.

\section{Observations} \label{sec:n3}

\begin{figure*}
\epsscale{1.8}
\vspace{-60pt}
\plotone{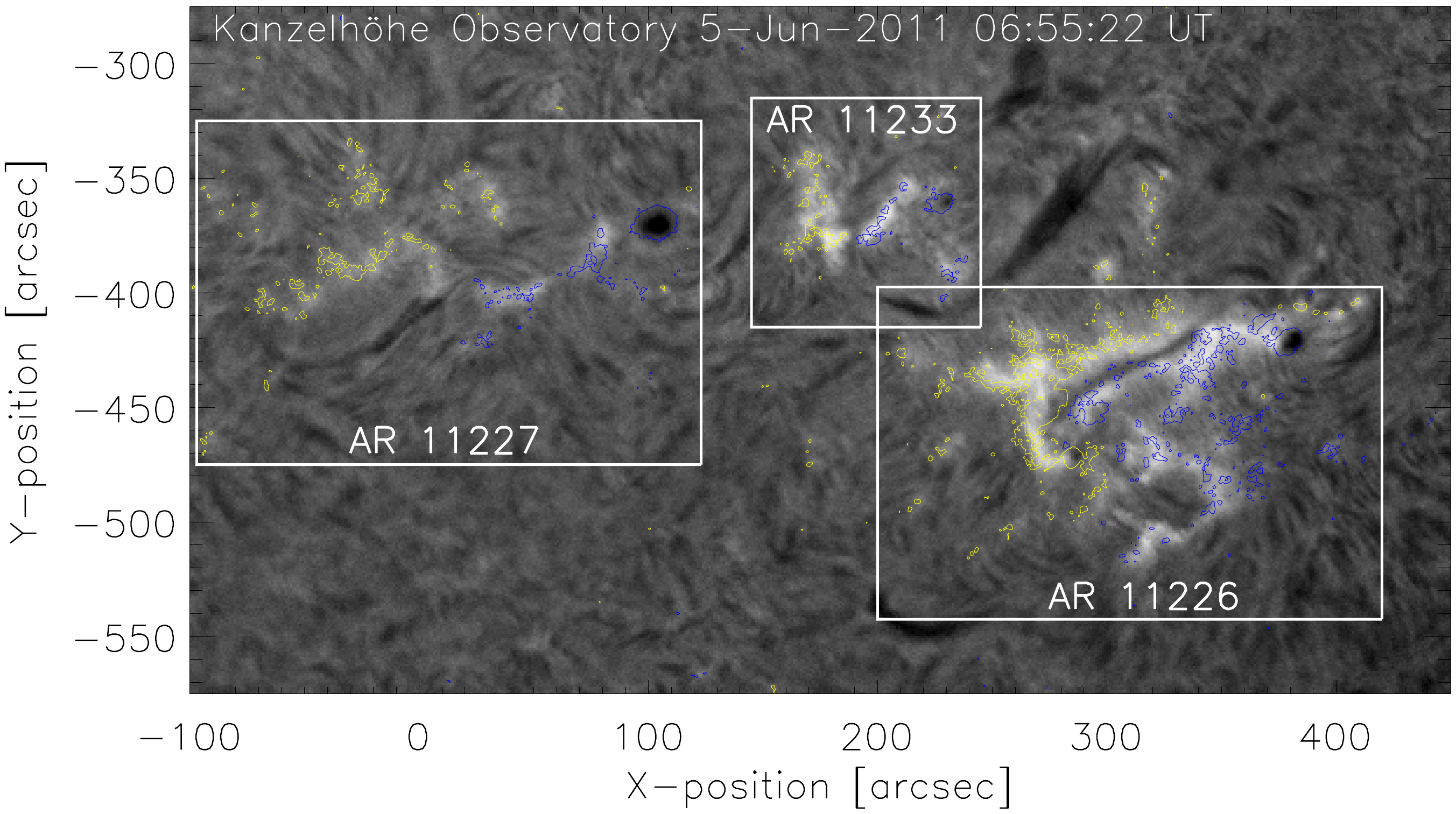}
\vspace{-10pt}
\caption{H$\alpha$ observation from the Kanzelh{\"o}he Observatory showing the three neighbouring ARs that form the active region complex. From east to west the ARs are: 11227, 11233 and 11226 \textbf{outlined by the white boxes}. The LoS magnetic field from HMI is shown in blue (yellow) contours representing positive (negative) polarities corresponding to levels of $\pm$300~G respectively. The image shows that filaments have formed along PILs within the ARs and between ARs. \label{fig1}}
\epsscale{1}
\end{figure*}

\begin{figure*}
\epsscale{2}
\vspace{-60pt}
\plotone{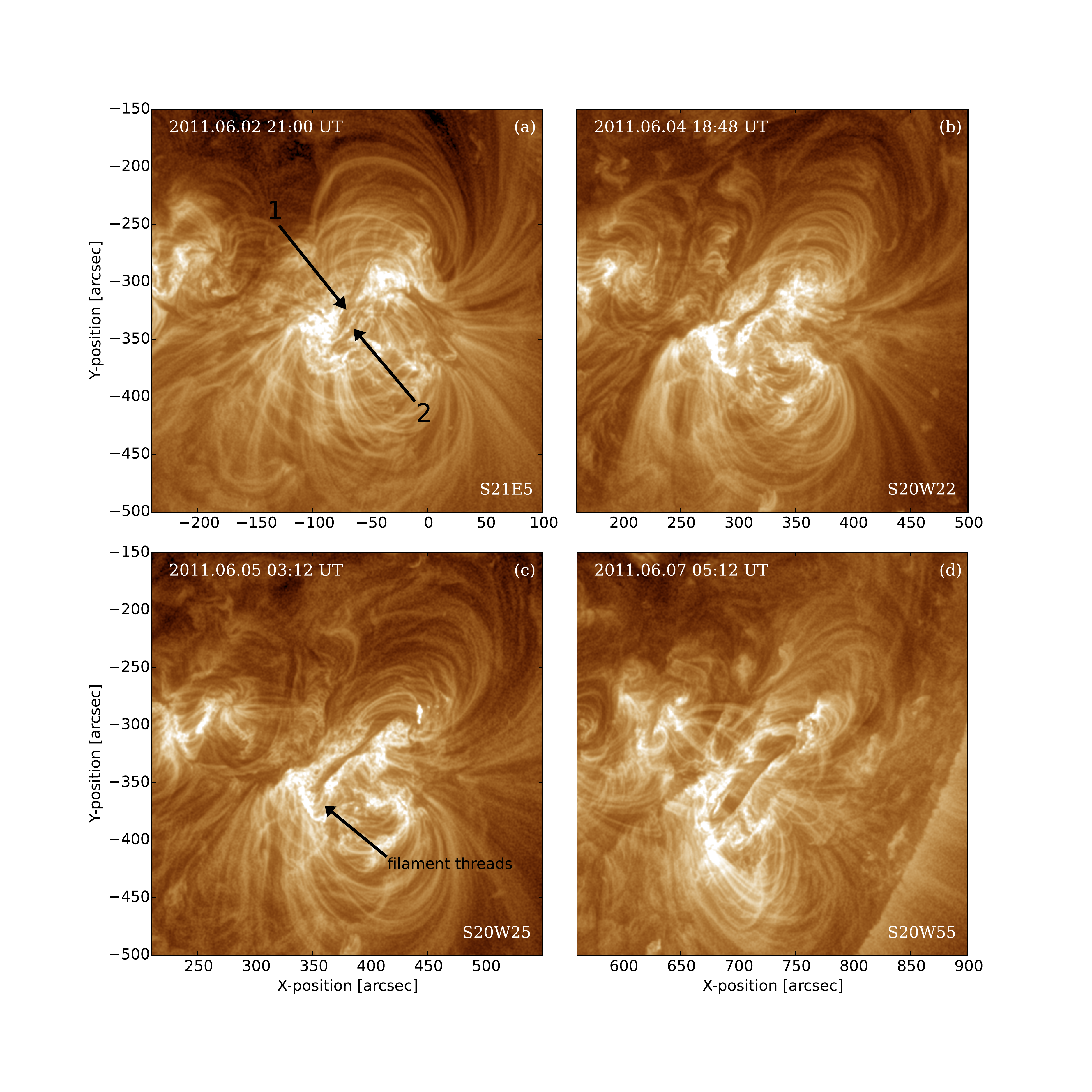}
\caption{Formation of the massive filament in AR 11226 observed in high-resolution 193~{\AA} images from SDO/AIA that have been processed using the Multiscale Gaussian Normalisation (MGN) technique of \citet{Morgan-2014}. The black arrows in panel (a) illustrate the two main sections of filament material that merge together to form one filament along the PIL. Dense filament threads that build into the filament during its evolution and extend the filament in length towards the south east are indicated by the arrow in panel (c). The extending filamentary threads in (c) are formed above the orphan penumbra shown in \textbf{(d)} and in Figure~\ref{fig4} on June 5. The location of AR centre is noted in the bottom right of each panel. \label{fig2}}
\epsscale{1}
\end{figure*}

\subsection{Filament Evolution}

The filament being studied here formed in AR 11226 which rotated onto the solar disk on 2011 May 27 and was part of a three AR complex; the other two regions being ARs 11227 and 11233, located to the east of AR 11226. See Figure~\ref{fig2} and also Figure~1 in \citet{Lidia-2014} for the arrangement of the ARs. Filament material is already present along the internal PIL of AR 11226 as it rotates over the limb. During its disk passage the filament is involved in two eruptions before the event that takes place on 2011 June 7. These two eruptions take place on 2011 May 29 and June 1. The CME on 2011 June 1 erupts from the region at  $\sim$16:00 UT. The eruption removes some, but not all, of the filament material. The filament therefore has a quiet phase between the CMEs on 2011 June 1 and June 7. 

Using the AIA 193~{\AA} waveband, the filament material that remains after the eruption on 2011 June 1 is seen to be present in two main sections, which overlap at the central part of the PIL (Figure~\ref{fig2} panel (a)). Late on June 2 the overlapping sections begin to thicken and have merged to form one long filament structure situated along the PIL by late on June 4 (Figure~\ref{fig2} (b)). Then, adjacent to the southern end of the filament, strands of relatively cool plasma build into the filament during June 5 and early on June 6 (Figure~\ref{fig2}(c)). The growth of the southern end of the filament appears to be mostly complete by noon on June 6 (Figure~\ref{fig2} (d)).

\begin{figure*}[t]
\epsscale{2.25}
\vspace{-60pt}
\plotone{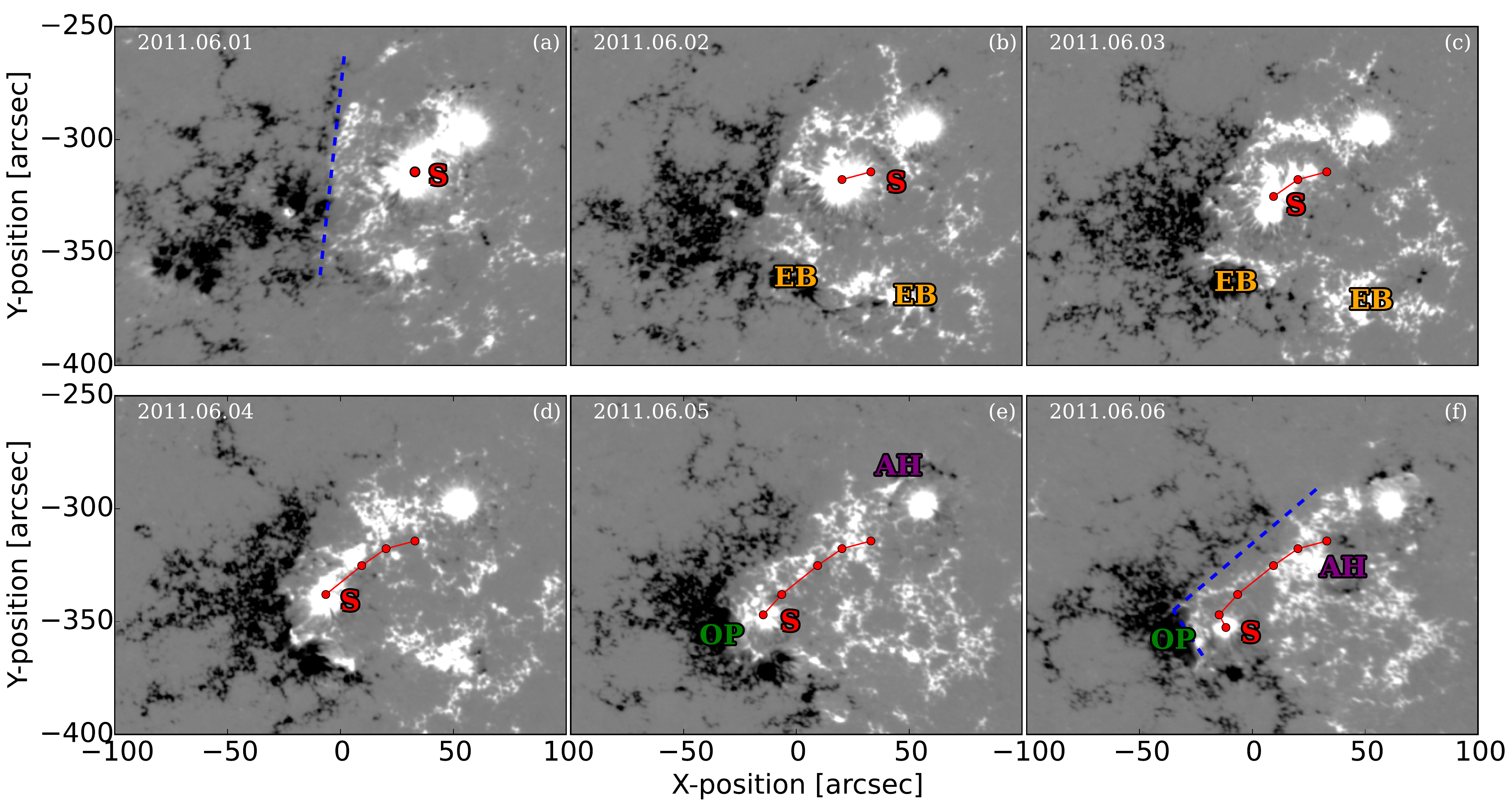}
\vspace{-10pt}
\caption{The line-of-sight photospheric magnetic field evolution of AR 11226 between June 1--6 as determined by HMI magnetograms displayed between $\pm$500~G at 00:00 UT and differentially rotated to disk centre (2011 June 3 04:10 UT). The following photospheric features that dominate the magnetic field evolution are highlighted in the following panels: (a) \& (f) show the C-shaped PIL and its change in inclination (blue dotted line), (a)-(f) show the rapid motion of the largest positive sunspot S, (b) \& (c) show the emerging bipole (EB) in the south and (e) \& (f) show the location of the two anti-Hale emerging bipoles (AH) around the stationary positive spot and the orphan penumbra (OP). The flux-weighted central coordinates of sunspot S that are calculated using the STEF algorithm are represented by the circle markers and solid line in red. \label{fig3}}
\epsscale{1}
\end{figure*}

\subsection{Photospheric Field Evolution}

The AR complex is studied during the period beginning on 2011 June 1 at 00:00~UT until June~6 06:00~UT (a day before the eruption being studied here). After June 6 06:00~UT, the leading edge of the positive polarity in AR 11226 is too close ($\sim$~60$^{\circ}$ longitude) to the limb to be able to make reliable flux measurements using the LoS magnetogram data. The three ARs in the complex differ in flux content. AR 11226 is a large region with a maximum flux of $8.2 \times 10^{21}$~Mx during the time period studied, whereas AR 11227 and 11233 are smaller regions with maximum fluxes of $3.2 \times 10^{21}$~Mx and $2.2 \times 10^{21}$~Mx, respectively. 

AR 11226 is seen to emerge on the far side of the Sun on 2011 May 25 in EUVI observations from STEREO B. From the Earth's perspective, AR 11226 rotates over the east limb in a roughly bipolar configuration. The opposite polarities are butted up against each other with initially the AR having umbrae of opposite polarity spots within a single penumbra. This is reflected in the Mount Wilson magnetic classification, which categorises the AR as having a $\beta \delta$ classification until May 30 followed by a $\beta \gamma$ configuration until June 4. The opposite polarities never fully separate and this suggests that AR 11226 is formed by the emergence of a  complicated configuration such as the emergence of a highly twisted or distorted flux tube.

Throughout the seven days leading up to the eruption, the photospheric field evolution of AR 11226 is very dynamic (Figure~\ref{fig3}). The positive polarity (leading) sunspot starts to split on 2011 May 31 and continues its division on June 1. One part remains as a stationary spot, whilst the largest section (labelled S in Figure~\ref{fig3}) breaks away and continues to move in the south-east direction towards the PIL. Then, around June 2 12:00~UT, the sunspot divides again into several portions with the largest (still labelled S) continuing to move rapidly towards the PIL. The proper motion of sunspot S was calculated by using the change in heliographic latitude and longitude in the Kanzelh{\"o}he data provided by the DPD catalogue. Over the period of three days (June 2--4) the velocity of the sunspot as it moves towards the PIL was calculated to be 0.19~km~s$^{-1}$.

Overall, the positive magnetic polarity becomes elongated and the sunspots are seen to exhibit substantial moving magnetic feature (MMF) activity, with a large proportion of these features streaming towards the PIL. In previous work it has been found that MMFs can play a role in filament evolution and eruption. \citet{Deng-2002} discovered that the activation and ejection of plasma ``blobs'' into a filament was a result of MMF cancellation involving features $\sim$10$^{19}$~Mx.

From June 1 to 4 there is a large addition of magnetic flux ($\sim$ $7 \times 10^{20}$~Mx up to $1.4 \times 10^{21}$~Mx) due to an emerging bipole located in the southern part of the AR (EB; Figure~\ref{fig2} (b) \& (c)). This bipole has the same magnetic orientation as AR 11226, which is of Hale orientation.

Two further episodes of flux emergence are seen to begin on June 4 22:00~UT and June 5 06:00~UT with an anti-Hale orientation (AH; Figure~\ref{fig2} (e) \& (f)). The bipoles emerge to the north and south of the stationary positive spot, respectively. The positive polarity of the second anti-Hale flux emergence is calculated, using the DPD catalogue, to have a velocity of 0.36~km~s$^{-1}$ in the south-eastward direction towards the PIL over a 13-hour period, mimicking the motion of spot S.

Strong cancellation is observed along the internal PIL, the orientation of which becomes increasingly tilted away from its initial N-S orientation with time as shown by the blue dashed line in Figure~\ref{fig2} (a) \& (f).

The unusual sunspot motions eventually cause the dispersed negative (following) polarity to become compressed as the process of cancellation reconnection is presumably not fast enough to remove the negative flux, causing a pile up at the south eastern end of the PIL. This is made apparent by the formation of strong field region (Figure~\ref{fig2} (e) \& (f), labelled OP). When viewed in the HMI continuum, the compression of the negative field is associated with the formation of an orphan penumbra, which later forms small spots at its periphery.

\subsection{Vector Field Evolution}

The evolution of the vector magnetic field has been studied during the same time period as the LoS field (2011 June 1 at 00:00~UT until June~6 06:00~UT). The top row of panels in Figure~\ref{fig4} show the evolution of the vector magnetic field at the southern end of the filament when the large sunspot S has collided with the negative polarity flux concentrations. As a result, opposite-polarity magnetic flux accumulates at the PIL and cancellation proceeds. During this time the transverse component of the vector field evolves to more strongly and coherently cross the PIL in the inverse direction (i.e. from negative to positive) at numerous locations below the filament. This is interpreted as observational evidence of the presence of bald patches (BPs), \textit{i.e.} of locations where the magnetic field is horizontal and forming a dip above the polarity inversion line where material can be sustained against gravity. The BP locations have been directly computed from the vector magnetogram data using Eq. 3 in \citet{Titov-1993}. Figure~\ref{fig4} shows the location of the BPs at four different times, indicating a coherent evolution of dips in time. Such a coherence is observed consistently in large sections along the PIL during the latter stages of the magnetic field evolution (from June 4 to June 6). Hence, it reduces drastically the possibility that the computed BPs are the random effect of an erroneous resolution of the 180-degree ambiguity. The concave-up magnetic field as deduced by the BP observations can in principle be produced by either the presence of a weakly twisted flux rope, or by the presence of an S-shaped PIL, see \textit{e.g.} Fig. 2 of \citet{Titov-1993}. In this case we have a PIL which is simple in shape: a C-shaped rather than an S-shaped PIL and so we interpret this as the presence of helical field in the form of a weakly twisted flux rope.

\subsection{Penumbra Formation}

Penumbra formation is characterised by the development of filamentary structures around a sunspot and is considered to indicate the presence of magnetic field that is close to horizontal. These structures form around pores above a certain diameter or magnetic flux content and are partial at first, appearing on the exterior of the spot away from the site of flux emergence. During the evolution of AR 11226 there are two significant episodes of unusual penumbra formation observed in the HMI continuum data. Firstly, there is the formation of partial penumbra in the negative polarity of the emerging bipole (EB) towards the interior of the bipole. This configuration is opposite to what is expected and is due to the bipole emerging in the vicinity of the negative polarity of AR 11226.

\begin{figure*}
\epsscale{2.25}
\vspace{-60pt}
\plotone{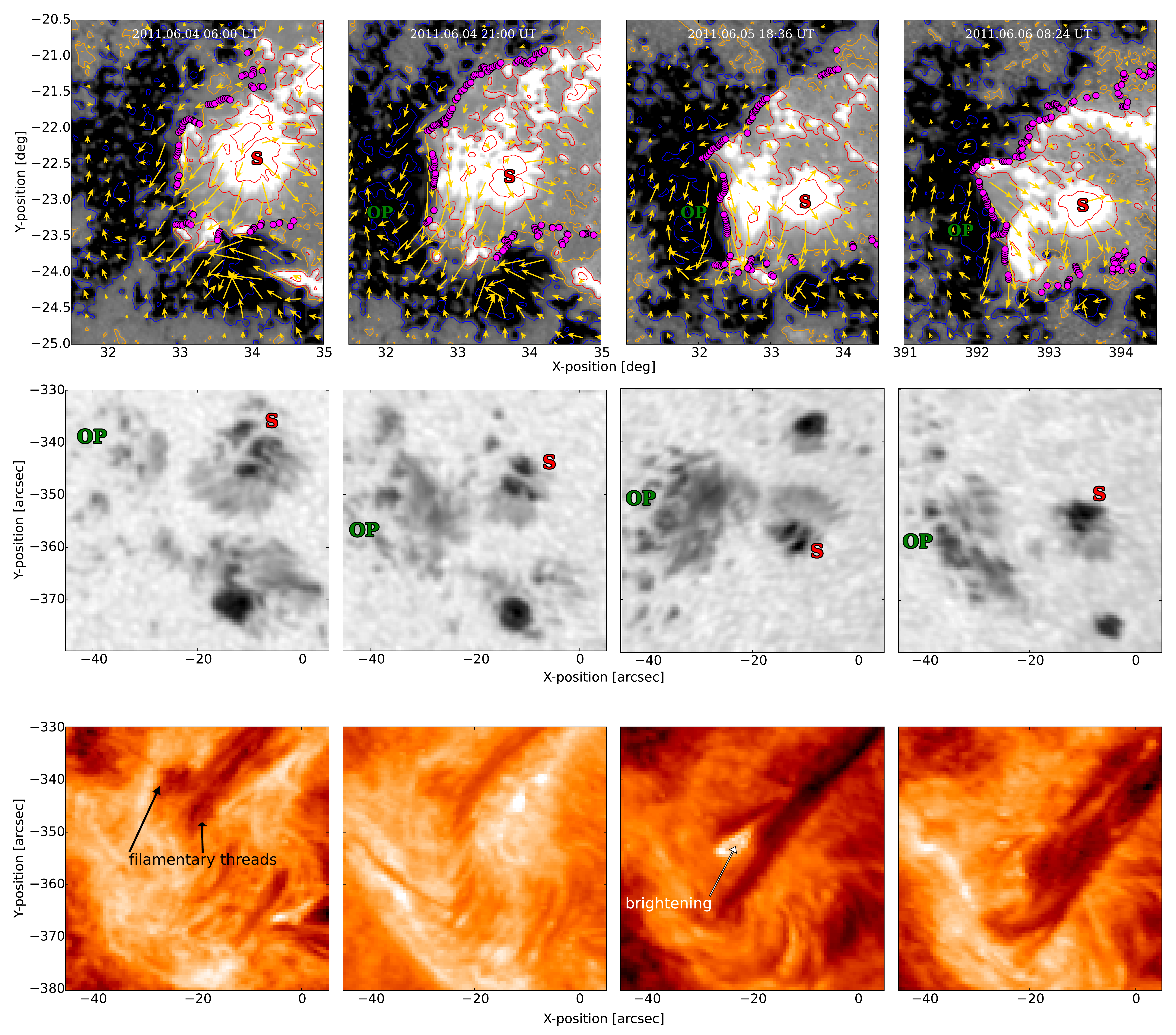}
\centering
\caption{The evolution of the HMI vector magnetic field (top), white light continuum (middle) and AIA 304~{\AA} data (bottom) between June 4--6. The transverse field, which is represented by the yellow arrows, is observed to inversely cross the PIL while the locations of the bald patches are represented by the points in magenta. The PIL and locations where the magnetic field has a value of 0~G is represented by the orange contours. The radial field is displayed between $\pm$~500~G with red and blue contours representing the positive and negative field respectively at values $\pm$~200 and 1000~G. The splitting and rapid motion of sunspot (labelled S) and the formation of the orphan penumbra (OP) is observed in the white light data. The arrows on the 304~{\AA} images show the presence and locations of filamentary threads (black) and the associated brightenings (white) when the threads build into the filament through magnetic reconnection. This causes the filament to grow and its footpoint to extend southwards. Movies of the HMI vector magnetic field [fig\_4a.mpg] and the continuum data [fig\_4b.mpg] are available in the online version. \label{fig4} }

\epsscale{1}
\end{figure*}

Secondly, an orphan penumbra forms in the location of negative polarity magnetic field, as mentioned above. The formation begins on June 5 around 23:00~UT when scattered penumbral areas develop, some of which include small spots, and some of which do not. The formation is apparently driven by the motion of sunspot S, which collides with and compresses the negative polarity at this time (Figure~\ref{fig4}). The orphan penumbra threads are located parallel to the PIL, above the negative polarity, directly to the east of spot S. The orphan penumbra lies underneath the southern end of the filament. As the filament grows (and extends further to the south) the orphan penumbra reduces in size and starts to disappear around June 6 06:00~UT. This occurs after the disappearance of a transient brightening late on June 5. These observations support the interpretation that the magnetic field is being reconfigured through magnetic reconnection, which could be responsible for injecting plasma into the magnetic field configuration supporting the filament.

\subsection{Flux Cancellation Rate}

\begin{figure}[h]
\plotone{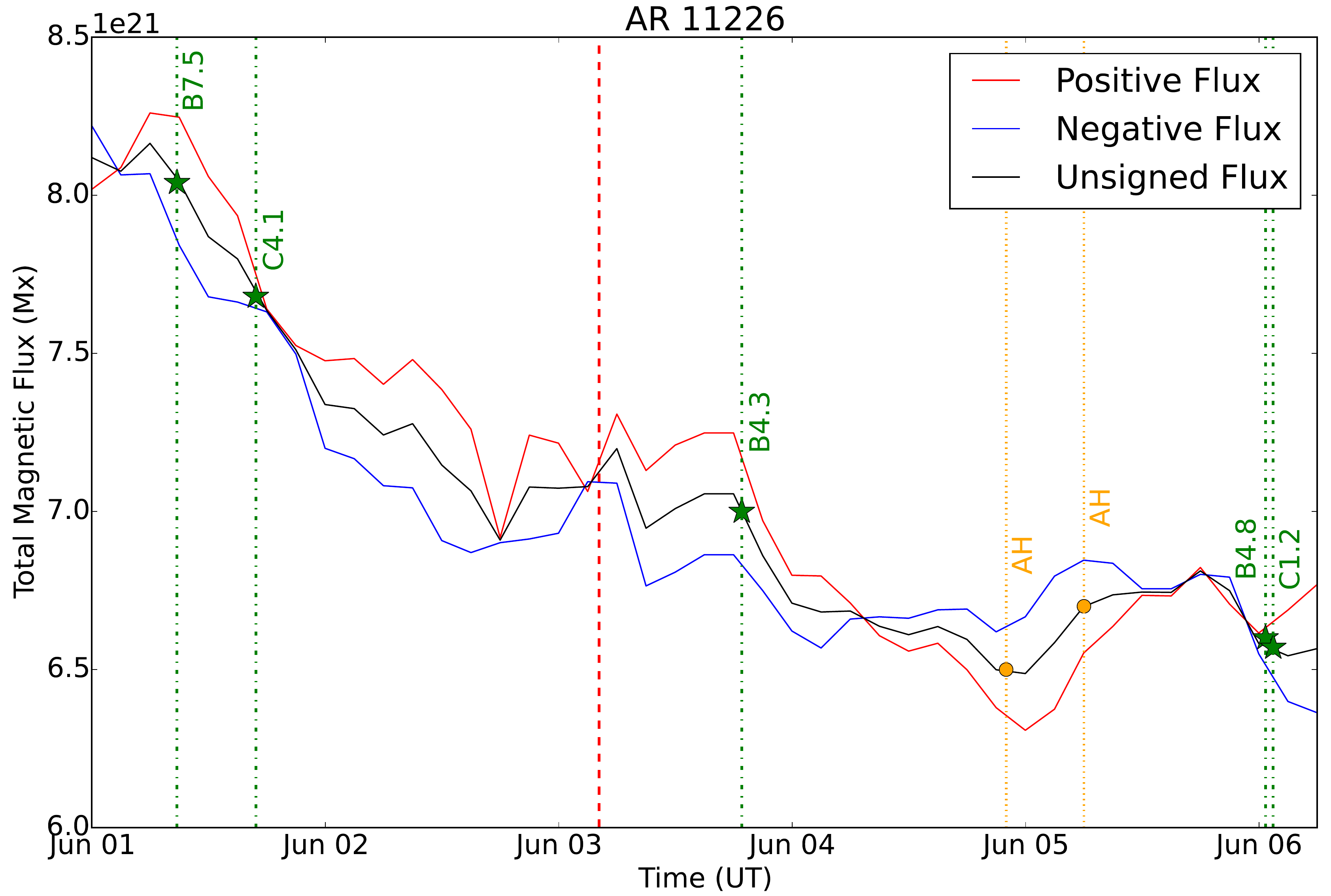}
\caption{Total positive (red), negative (blue) and unsigned (black) magnetic flux for AR 11226, determined from the HMI data using the STEF algorithm, over the six-day period beginning on June 1. The dashed red line indicates the point at which the AR passed central meridian. The green dashed dot lines and stars represent the timings of solar flares and their corresponding GOES class. The orange points and dashed lines indicate the start of the emergence of the two anti-Hale bipoles (AH). \label{fig5}}
\end{figure}

In order to determine the flux cancellation rate in AR 11226, the magnetic flux from the bipole that emerges in the southern part of the AR must be taken into account, as this introduces additional magnetic field to the region and masks the true cancellation rate. Because of this, the flux of the emerging bipole was measured so that it could be subtracted from the total AR flux (which contains both the pre-existing and emerging magnetic field). However, this becomes challenging, especially on June 3, when it is very hard to separate the flux of the emerging bipole from the surrounding AR magnetic field. Due to this, the measured magnetic flux of the bipole was subtracted across the first three days (June 1 00:00~UT -- June 3 15:00~UT) of the time interval over which the entirety of the AR flux is measured.  For the remainder of the time range the final flux measurement of the bipole was subtracted. The flux evolution of the AR, with that of the emerged bipole subtracted, is shown in Figure~\ref{fig5}. This revealed large amounts of flux cancellation in the region that was viewed in the observations but originally masked by flux emergence in the flux measurements.

After a slight initial increase in magnetic flux, there is roughly a 3.5 day period (2011 June 1 09:00~UT to 5 00:00~UT) when the unsigned flux of the AR is decreasing due to ongoing flux cancellation.  The majority of the flux cancellation is occurring at the internal PIL with an average flux cancellation rate of $2.0 \times 10^{19}$~Mx~hr$^{-1}$. In total, $1.7 \times 10^{21}$~Mx is cancelled during this time period. The flux in the AR at the start of the period studied is $8.2 \times 10^{21}$~Mx and so the amount of flux cancelled represents 21\% of the AR flux. Following the approach of \citet{Ballegooijen-1989} and \citet{Green-2011}, an amount of flux equal to that cancelled is available to be built into the magnetic configuration that contains the filament material.

\begin{figure}[h!]
\plotone{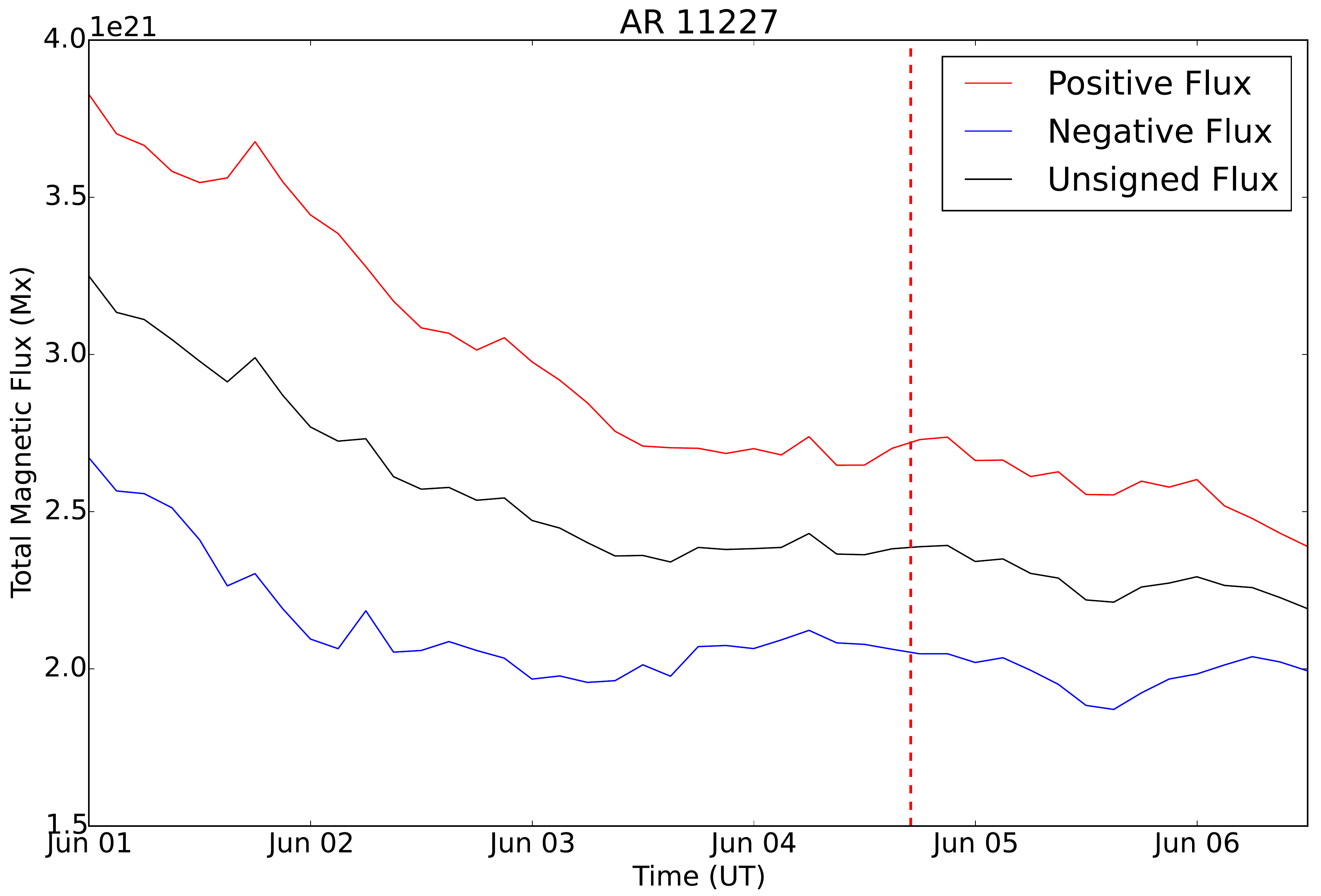}
\caption{Total positive (red), negative (blue) and unsigned (black) magnetic flux for AR 11227 over the same period as Figure~\ref{fig5}. The dashed red line indicates the point at which the AR passed central meridian. \label{fig6}}
\end{figure}

\begin{figure}[h!]
\plotone{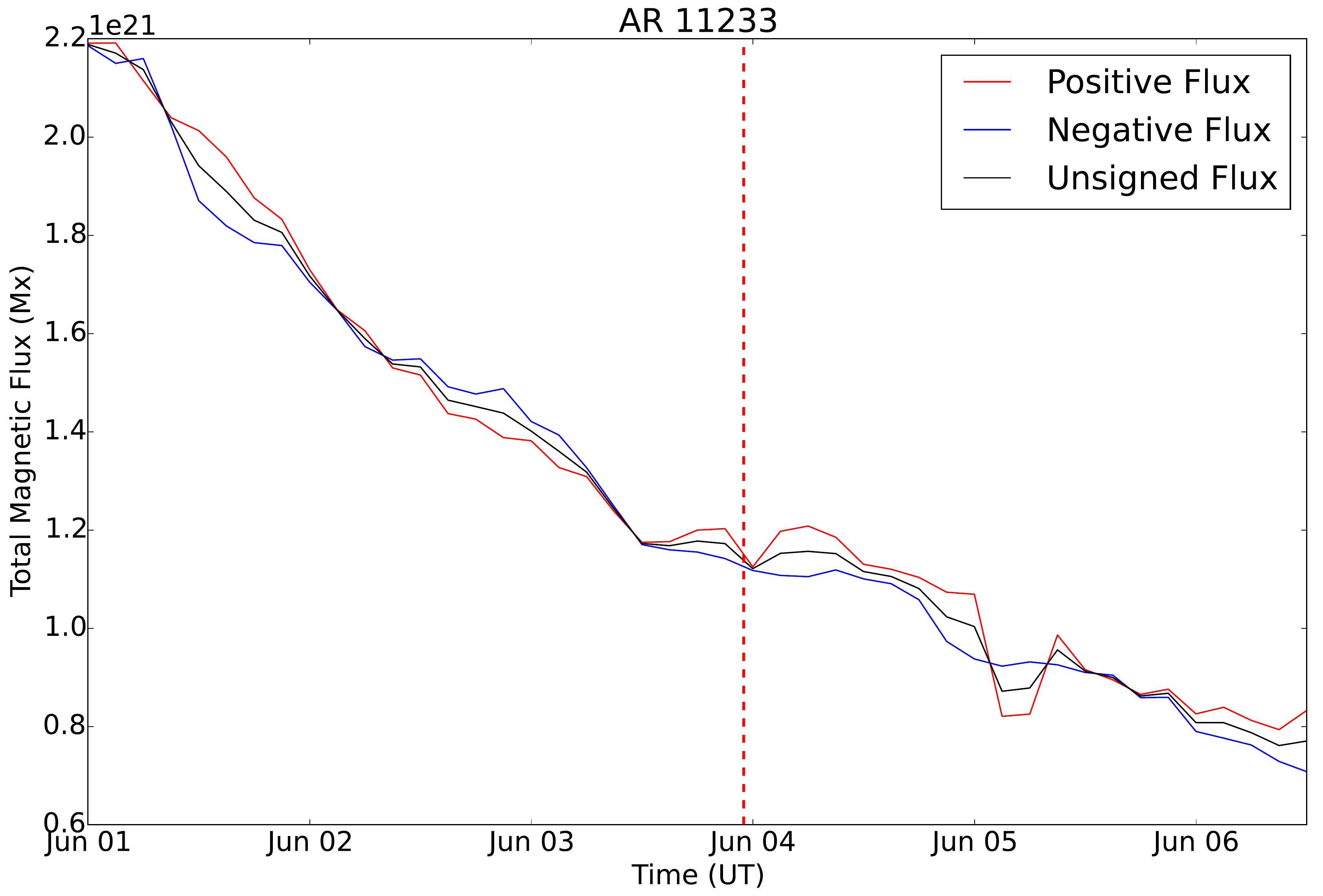}
\caption{Total positive (red), negative (blue) and unsigned (black) magnetic flux for AR 11233 over the same period as Figure~\ref{fig5}. The dashed red line indicates the point at which the AR passed central meridian. \label{fig7}}
\end{figure}

In contrast, the flux evolution of the two neighbouring ARs 11227 and 11233 (Figures~\ref{fig6} \& \ref{fig7}) show lower flux cancellation rates. Hence, they have lower flux cancellation values over the time period studied: AR 11227 loses $1.0 \times 10^{21}$~Mx of flux, which represents 30\% of the peak value during the study period, and AR 11233 loses $1.2 \times 10^{21}$~Mx of flux, which is 54\% of the maximum value (as shown in Table~\ref{table1}). Most of the flux cancellation in AR 11226 occurs along the internal PIL; by contrast, the other regions display cancellation mainly at their peripheries, therefore building filaments between adjacent ARs (Figure~\ref{fig1}).

The total photospheric magnetic flux of AR 11226 can be interpreted as having two main phases of evolution. There is a period of ongoing flux cancellation at the internal PIL due to the polarities being ``butted up" against each other that is aided by the motion of spot S, bringing significant quantities of positive flux to the PIL. During this period, the two filament sections observed in the AIA 193~{\AA} merge into one. Sunspot S then becomes stationary when it reaches the negative polarity. In this phase, the orphan penumbra and new ``filament threads" are observed. These threads build into the main filament, extending its southern end. Since we observe brightenings in the threads, it is interpreted that the facilitating process is magnetic reconnection. However, it is during this phase that two anti-Hale bipoles are seen to emerge in the vicinity of the stationary leading spot, masking flux cancellation during the second phase. The flux in the anti-Hale bipoles is not removed from the flux evolution plot of AR 11226 because of the difficulties in distinguishing it from the background field of the AR.

\section{Discussion} \label{sec:n4}

\subsection{Photospheric Motions \& Global Field Structure of AR 11226}

The splitting of the positive spot in AR 11226, and the rapid motion of the sections that break away, is highly unusual and rarely observed. This complex and dynamic nature of the magnetic field indicates that the flux tubes from which the AR forms may have differing sub-surface configurations. One of the possibilities is that a U-loop emergence is driving the opposite polarity magnetic fields towards one another (see e.g. \citealt{Lidia-2000}). However, for the AR studied here we suggest an alternative interpretation. Initially, the AR is bipolar overall with a Hale orientation, supporting the interpretation that AR 11226 is formed from the rise of an $\Omega$-loop originating from the toroidal field at the base of the convection zone. The other two ARs in the three-region complex and the bipole that emerges at the southern edge of AR 11226 also have a Hale orientation.

The two bipoles that emerge later in the evolution of AR 11226 on June 5 and 6 have an anti-Hale orientation (AH in Figure~\ref{fig2}), possibly produced as a result of sub-surface vortices or the kink instability acting on the sub-photospheric flux tube \citep{Lopez-2003}. We speculate that even the splitting of the positive leading spot on May 31, and the subsequent fast motion of spot S (Figure~\ref{fig2}) towards the PIL, is also due to the emergence of an anti-Hale orientated flux tube and its sub-surface interaction with the pre-existing Hale-orientated flux system. This is analogous to the flux emergence and sub-surface cancellation example described by \citet{Wang-1993}. In both cases, the new flux emergence takes place in strong monopolar field. Sub-surface reconnection prevents the oppositely orientated polarity from being observed. In this case, the negative polarity of the anti-Hale bipole remains mostly hidden below the photosphere. The reorganisation only reconnects part of the positive spot's magnetic field to the buoyant emerging flux. The sub-surface connections of the part which becomes spot S change, and this reconfiguration causes the splitting and motion of this section of the positive polarity, whilst the rest remains stationary. The spot moves south-eastward towards the PIL as the emerging anti-Hale flux tube's consecutive cross-sections with the photosphere are shifting south-eastward.

In this AR we interpret the ongoing flux cancellation along the PIL, and associated magnetic reconnection, to lead to the formation of two magnetic flux systems. A small loop that cannot always be resolved against the surrounding AR field and which submerges through the photosphere (observed as a reduction in magnetic flux) and a concave-up section of field (BP locations), which is observed as an inverse crossing of the PIL at the photosphere. This interpretation along with the combined observations of the photospheric magnetic field, EUV emission and absorption structures are well aligned to the evolutionary sequence of filament and flux rope formation as laid out in \citet{Ballegooijen-1989}.

The alternative explanation of the BPs present along the PIL as a dipped arcade-like field lines touching over an S-shaped PIL is difficult to construct in our case without invoking helical field. The problem lies in the fact that, in this case, we do not have an S-shaped PIL as in \citet{Titov-1993}, but rather a simple C-shaped PIL. Hence, such a dipped-arcade like field line should, for example, start from the positive polarity in the northern part of the magnetogram, pass above the PIL, turn and touch the PIL with an inverse crossing, then again turn back and cross high up over the PIL to finally be rooted in the negative polarity. Such a field line, if not helical, would be highly non-force-free, which is difficult to justify in the corona. Furthermore, there is no indication of the presence of such field lines in Figure~\ref{fig4}. From the evidence outlined above, we are therefore compelled to infer the presence of a weakly twisted flux rope as the single magnetic structure that can explain several observations. Moreover, the subsequent eruption was modelled by \citet{Lidia-2014} using a flux rope located at the same position of the observed filament. The numerical simulation of the magneto-hydrodynamical evolution of the magnetic field was shown to capture many of the essential features of the observed eruption, to a very high degree of accuracy, in this way corroborating our hypothesis of flux rope formation.

As the breakaway spot S moves towards the PIL and compresses the pre-existing field in the filament channel, an orphan penumbra forms. Previous formation mechanisms of penumbrae have included their formation as the result of a flux rope trapped in the photosphere \citep{Kuckein-2012a, Kuckein-2012b}, the emergence of an $\Omega$-loop trapped by canopy fields \citep{Lim-2013, Zuccarello-2014} or submerging horizontal fields \citep{Jurcak-2014}. Observational evidence in this study suggests the existence of a low-lying flux rope (and filament) with highly sheared horizontal field at the PIL. However, in this case there is no emergence of magnetic flux in the region forming the orphan penumbra during this period ($\sim$June 4--6), which suggests that the compression of the negative magnetic field is responsible.

Filament threads that consist of relatively cool plasma are seen to form in 304~{\AA} data (Figure~\ref{fig4}), connecting spot S and the negative polarity of the newly emerging anti-Hale bipole. This allows plasma to be injected into the filament through magnetic reconnection, which is evident through transient brightenings. Previous observations by \citet{Zhang-2014} have described this ``flux-feeding'' process where chromospheric fibrils feed flux into the filament eventually leading it to become unstable to the torus instability. \citet{Wang-2013} observed large brightenings in the regions where flux cancellation is occurring along the filament channel and suggest that the associated loop systems go on to form part of the filament channel. Another study by \citet{Liu-2012} has found similar evidence whereby flux is transferred from a lower to an upper branch in a ``double-decker" filament configuration. During this phase the absorption and extent of the filament increases, which is interpreted as mass being transferred into the filament (known as ``mass loading"). This occurs in the evolutionary phase of the filament when it is close to eruption. This process of mass accumulation may force the horizontal field to a lower height, raising the possibility that magnetic reconnection associated with flux cancellation is occurring very low in the photosphere. 

\subsection{Flux Cancellation}

The average flux cancellation rate along the PIL where the filament is located in AR 11226 is $2.0 \times 10^{19}$~Mx~hr$^{-1}$. In total, $1.7 \times 10^{21}$~Mx is cancelled during the time period studied. This rate exceeds that found in most of the previous studies (see Table~\ref{table1}). \citet{Vemareddy-2014} records a higher value, but this is omitted from Table \ref{table1} as the size of the region in that work is comparable to the entire June 7 three AR complex with multiple PILs. \citet{Sterling-2010} also record a higher cancellation rate ($4.2 \times 10^{19}$~Mx~hr$^{-1}$) compared to what we find for AR 11226 ($2.0 \times 10^{19}$~Mx~hr$^{-1}$), but over a shorter time period of one day. AR 11226 exhibits the largest amount of flux cancellation yet studied, over the longest period of time.

The data produced by the HMI instrument are of high quality, however there are known uncertainties, limitations and systematic errors present that affect the measurement of magnetic flux. These include a sinusoidal variation in the total magnetic flux with a periodicity of 12 and 24 hours, due to a Doppler shift present in the Fe spectral line \citep{Hoeksema-2014}. The main contribution to this shift is the geosynchronous orbit of SDO. The $\pm$3~km~s\textsuperscript{-1} daily variation in spacecraft orbital velocity causes a sinusoidal variation in the total flux measured. This affects weak and strong magnetic field strengths in the LoS magnetograms differently, with the daily variation remaining below 30~G for field strengths below 1000~G and less than 75~G for field strengths below 2250~G. On average during a day this is roughly $\pm$~35~G \citep{Couvidat-2015}. Nevertheless, the strength of these instrumental and observational effects do not account for the strong flux cancellation we observe in AR 11226 prior to eruption. The leading edge of the positive polarity of AR 11226 reaches $\sim$60$^{\circ}$ from central meridian on June 6. Due to the spatially dependent sensitivity of HMI the noise level increases as a function of the centre-to-limb angle and the spacecraft's orbital velocity. This increases the value of low and moderate field pixel values (between 250--750~G) by a few tens of percent and manifests itself as broad peaks that are centred at $\sim \pm$60$^{\circ}$ in the magnetic flux. This could be responsible for the increase in flux seen in Figure~\ref{fig5} at this time. Although, the increase also coincides with the emergence of the two anti-Hale bipoles and so it is difficult to disentangle these effects. The proximity of AR 11226 to the limb on June 6 means that the magnetic flux was not measured right up until the time of the eruption but ceased approximately a day beforehand. Therefore, these results are a lower limit of the total flux cancelled in the lead up to the eruption. Furthermore, due to the fact cancellation persists for several days, we can extrapolate that if the cancellation process continues at the same rate ($2.0 \times 10^{19}$~Mx~hr$^{-1}$) as we observed in the period of June 1 09:00~UT to June 5 00:00~UT up until eruption, an extra $1.1 \times 10^{21}$~Mx could have been cancelled. This would therefore result in a total amount of $2.8 \times 10^{21}$~Mx flux cancelled between two consecutive CMEs.

\section{Conclusions} \label{sec:n5}

This work studies the formation of the exceptionally massive filament in AR 11226, which erupted on 2011~June~7 at $\sim$06:15~UT. We suggest that magnetic reconnection associated with strong flux cancellation was responsible for the formation of the filament and the injection of plasma into a forming flux rope. This was evident from the following photospheric and chromospheric signatures:

\begin{itemize}

	\item The presence and coherent appearance of bald patches in the transverse magnetic field component along the PIL. This is interpreted as the observational evidence of concave-up magnetic field geometry or magnetic dips in the presence of a weakly twisted flux rope.
    
	\item A high flux cancellation rate ($2.0 \times 10^{19}$~Mx~hr$^{-1}$), which is driven by the motion of the spot S towards the PIL.
	
	\item The formation of an orphan penumbra and small spots as the dispersed negative magnetic field is compressed by the positive-polarity spot S reaching the PIL. We suggest that the process of cancellation reconnection is not fast enough causing a pile up of negative flux at the south eastern end of the filament and PIL. 
	
	\item The south eastern end of the forming filament is observed to be rooted in the location of the orphan penumbra. The penumbral threads lie parallel to the PIL. In the chromosphere in the same region we observe highly sheared filamentary threads. These dense threads are reconfigured through the process of magnetic reconnection, extending the filament towards the south east and possibly injecting plasma into the filament.
	
\end{itemize}

The flux cancellation that occurred in the three ARs was computed by applying the STEF algorithm to calculate the reduction in unsigned magnetic flux, roughly over a 3.5 day period (June 1 09:00~UT -- 5 00:00~UT). In AR 11226 $1.7 \times 10^{21}$ Mx (21\% of the maximum AR flux), equivalent to a small AR is cancelled up to a day before the eruption. This corresponds to a large cancellation rate of $2.0 \times10^{19}$~Mx~hr$^{-1}$ over a prolonged period, among one of the largest rates of cancellation of any previous studies to date. Assuming that this cancellation rate continued through the time period when HMI data cannot be used, due to instrumental and observational effects close to the limb, leads to an estimated $2.8 \times 10^{21}$~Mx of flux cancelled prior to the large filament eruption and CME on 2011 June 7.

The majority of the flux cancellation in AR 11226 occurred at the internal PIL, at the location where the filament formed. Whereas, for the neighbouring ARs the cancellation occurred at both the internal and external PILs, hence forming inter-active region filaments. The amount of flux cancelled along the internal PIL in AR 11226 is substantially more than the cancelled flux of the neighbouring regions during the same period; AR 11227 ($1.0 \times 10^{21}$~Mx) and 11233 ($1.2 \times 10^{21}$~Mx). 

The flux cancellation along the internal PIL of AR 11226 can be used to investigate the amount of flux that was built into the flux rope that was identified using the line-of-sight magnetic field data. However, the amount of flux cancellation observed may differ from the amount of flux that is built into the rope. By modelling weakly twisted flux ropes and validating against observations, \citet{Savcheva-2012} found that their modelled flux ropes contain around 60 to 70\% of the flux that is cancelled in the region in which the ropes form. Applying this conclusion to AR 11226, we infer that a substantial amount of flux was built into the flux rope that contained and supported the filament. Using the estimated value of the amount of flux cancelled in AR 11226 from the start of the study up to the point of eruption (both the quantity observed and the estimated quantity using the cancellation rate) leads us to estimate that between $1.7 \times 10^{21}$~Mx and $2.0 \times 10^{21}$~Mx of flux is in the rope when the filament eruption occurs. This is possibly a lower estimate because there was already a filament (and inferred partially formed flux rope) present at the start of our flux cancellation study. In any case, this quantity is around a factor of three higher than the flux ropes modelled by \citet{Savcheva-2012} which were formed in small and decaying bipolar ARs. This is 34~\% of the maximum AR flux ($8.2 \times 10^{21}$~Mx) that the AR contained at the start of the study. It is only at this point, when the ratio of the flux contained in the rope to the flux of the overlying arcade field is potentially 1:0.9, that the rope with its huge quantity of filament mass, can finally no longer be held down by the overlying arcade field. This would explain the development of the unusually massive filament and spectacular eruption of June 7.

\acknowledgments

The authors are thankful to the \textit{SDO}/ HMI and AIA consortia for the data, and the use of JHelioviewer (\url{http://jhelioviewer.org/}) for browsing data. We are grateful to Todd Hoeksema for the helpful discussions regarding the HMI data and to Dave Long for providing processed AIA images. We are also grateful to Pascal D{\'e}moulin for his valuable comments and would like to thank the SunPy community and David Perez-Suarez for SunPy guidance. We acknowledge the use of H$\alpha$ data provided by the Kanzelh{\"o}he Observatory. SLY and SD acknowledge STFC for support via PhD studenship. LMG is grateful to the Royal Society for a University Research Fellowship. LMG, LvDG and GV acknowledge the support of the Leverhulme Trust via the grant RPG-2014-05 ``Solar magnetic activity: bridging the gap between observation and theory." LvDG’s work was supported by the STFC Consolidated Grant ST/H00260X/1 and the Hungarian Research grant OTKA K-109276.

\begin{deluxetable}{ccccc} 
\tabletypesize{\scriptsize} 

\tablewidth{0pt}
\tablecolumns{5} 
\tablecaption{Flux cancellation along polarity inversion lines associated with filaments. \label{table1}} 
\tablehead{ 
& \colhead{Maximum Flux} & \colhead{Total Flux Cancelled} & \colhead{Duration}  & \colhead{Flux Cancellation Rate} \vspace{-0.2cm} \\ 
\colhead{AR} &\colhead{} & \colhead{} & \colhead{} & \colhead{} \\
 \colhead{} & \colhead{10$^{21}$~Mx}  & \colhead{10$^{21}$~Mx} & \colhead{Days} & \colhead{10$^{19}$~Mx~hr$^{-1}$}}
 
\startdata 
11226 & 8.2 & 1.7 (2.8) & 3.6 (5.9) & 2.0\\
11227 & 3.3 & 1 & 3.5 & 1.0 \\
11233 & 2.2 & 1.2 & 3.5 & 1.2 \\
$2007/10^{a}$ & 3.2 & 1.0 & 4.0 & 1.0\\
$10977^{b}$ & 2.1 & 0.71 & 2.5 & 1.2 \\
$10956^{c}$ & 10.0 & 1.0 & 1.0 & 4.2 \\
$1984/08^{d}$ & - & 0.5 & 5.0 & 0.5 \\
\enddata 

\tablecomments{$^{a}$\citet{Baker-2012} measured the peak positive AR flux leading up to an eruption. 
$^{b}$\citet{Green-2011} studied the flux evolution of both polarities, recording cancellation of negative flux during flux rope formation, leading to an eruption.
$^{c}$\citet{Sterling-2010} follows the total flux evolution over a six day period, recording total flux cancellation over two days prior to eruption.
$^{d}$\citet{Martin-1985} do not record total flux, however a flux cancellation rate is recorded over the AR decay period leading up to an eruption. 
}
\end{deluxetable}

\clearpage

\end{document}